\documentclass[11pt]{article}
\usepackage{a4wide}
\usepackage{amsmath}
\usepackage{amsfonts}
\usepackage{amssymb}

\newcommand{\ba}{\begin{equation}}
\newcommand{\ea}{\end{equation}}
\usepackage{graphicx}

\def\one{\mathbf{1}}
\def\oneP#1{\one_{\tau}}

\newtheorem{prop}{Proposition}

\newtheorem{coro}{Corollary}
\newtheorem{defi}{Definition}

\newtheorem{note}{Note}

\begin{document}

\author{Cyril Grunspan\footnote{I wish to thank Yann Braouezec, Rama Cont, Daniel Gabay, Martino Grasselli and Julien Warmberg for their valuable remarks on this short paper. Needless to say,  the usual disclaimers apply.}\\
ESILV,  Department of Financial Engineering\\
92916 Paris La D\'efense Cedex \\
cyril.grunspan@devinci.fr} 

\title{ {A Note on the Equivalence between the Normal and the Lognormal Implied Volatility : A Model Free Approach }}

\maketitle

\begin{abstract}
    \noindent First, we show that implied normal volatility is intimately linked with
    the incomplete Gamma function. Then, we deduce an expansion on implied normal volatility in terms of 
    the time-value of a European call option. Then, we formulate an equivalence between the implied normal
    volatility and the lognormal implied volatility with any strike and any
    model. This generalizes a known result for the SABR model. Finally, we
    adress the issue of the ``breakeven move'' of a delta-hedged portfolio.
\end{abstract}

\noindent JEL Classification G12 G13 C65 \bigskip


\section{Introduction}
	This article comes within the scope of the study of the asymptotics of implied
  volatility which has been considered extensively (\cite{RR}, \cite{GaoLee}, \cite{costeanu}). 
  Asymptotics of implied volatility are important for different
  reasons. First, they give information on the behaviour of the underlying
  through the moment formula (\cite{lee}) or the tail-wing formula (\cite{BenaimFritz1}). 
  Second, they allow a full correspondence between vanilla prices
  and implied volatilities. With such a correspondence, asymptotics in call
  prices can be easily transformed into asymptotics in implied volatilities.
  When applied to a specific model, asymptotics are widely used as smile
  generators (\cite{hagan}). In practice, other models are then used for
  pricing options using tools like Monte-Carlo simulations.
  
  \medskip
  
  So far, all the asymptotics studied by authors concern asymptotics for implied
  lognormal volatility. In this article, we consider implied normal
  volatility which refers to the Bachelier model. Why is it interesting to consider normal implied volatility?
  First, for short maturities, the Bachelier process makes more sense than the Black-Scholes model. Indeed,
  the behaviour of the underlying from one day to another is generally well
  approximated by a Gaussian random variable (see \cite{schachermayer}). 
  That's the reason why the Bachelier model is very popular in high frequency 
  trading (\cite{avella}). Second, the ``breakeven move'' of a delta-hedged portfolio option is easily interpreted as  
  normal   volatility. Generally, the P \& L of a book of delta-hedged 
  options is positive if the (historical) volatility of the
  underlying is greater than a breakeven volatility which has to be expressed
  in normal volatility. Moreover, it makes more sense to compare
  implied normal volatilities with historical moves of the underlying as can
  be done by a market risk department. Likewise, some markets such as fixed-income markets with
  products like spread-options are quoted in terms of implied normal volatility (\cite{kienitz}). 
  Finally, the skewness of swaption prices is much reduced if priced in terms of 
  normal volatility instead of lognormal volatility. Therefore, it is important to have
  a robust and quick way to compute implied normal volatilities from market
  prices and also to be able to switch between lognormal volatilities and
  normal volatilities. 

  \medskip
  
  What kind of asymptotics should we consider? Most of the approximations in option pricing theory are made under the
  assumption that the maturity is either small (see the Hagan et al formula \cite{hagan} for
  instance) or large (\cite{fouque}); it is actually assumed that a certain time-variance $\sigma^2 T$ 
  is either small or large. A possible way to derive such approximations is to replace the
  factor of volatility $\sigma$ by
  $\varepsilon \sigma$ and then set
  $\varepsilon=1$. This can be done at the partial differential
  equation level (see all the techniques coming from physics \cite{hagan}) as well as directly at the
  stochastic differential equation level with the help of the Wiener chaos theory for instance \cite{osajima}. 
  Other types of asymptotics are obtained by considering large strikes. In our approach, we
  unify all these types of asymptotics (see \cite{GaoLee} and \cite{grunspan} for the lognormal
  case). Indeed, we obtain an approximation of the implied normal volatility
  as an asymptotic expansion in a parameter $\lambda$ for $\lambda\ll 1$ and it turns out that 
  $\lambda\rightarrow 0$ when $T\rightarrow 0$ or $K\rightarrow +\infty$.

  \medskip
  
  This study is organized as follows. We first give another expression for the
  pricing of a European call option which involves an incomplete Gamma
  function (Proposition \ref{TVGamma}). Then, we inverse this function asymptotically and
  obtain an expansion of normal implied volatility. This is particularly
  important if we want to quickly obtain the implied normal volatilities from
  call prices as is the case in high frequency trading (\cite{avella}). The formula
  is also potentially useful theoretically if, given an approximation
  for the price of a European call option or a spread option (for instance in the framework of the Heston or the SABR
  model), we want to obtain an approximation of the normal implied volatility. 
  Finally, we restrict our formula to the order $0$ and we compare it to a similar formula
  for the lognormal case. Then, we obtain an equivalence between normal volatility and lognormal
  volatility. We use it also to compare the Black-Scholes greeks to the Bachelier greeks. Finally, we consider 
  a delta-hedged portfolio and we compute the ``breakeven move'' in the normal case as well as in the lognormal case.
  
\section{Another pricing formula for call options in the Bachelier model}
In a Bachelier model, the dynamic of a stock $(S_t)$ is given by:
$$
d S_t = \sigma_N  d W_t,
$$

with initial value $S$ at $t = 0$. The so-called "normal volatility" $\sigma_N$ is related to the price of a call
$C(T , K)$ struck at $K$ with maturity $T$ by the formula \cite{schachermayer}:

\ba
\label{BSN}
C(T , K) = (S - K) N\left(\frac{S-K}{\sigma_N \sqrt{T} }\right) + \sigma_N \sqrt{T} n\left(\frac{S-K}{\sigma_N \sqrt{T}}\right)
\ea

with
$$
n(x) =\frac{1}{\sqrt{2 \pi}}  \exp \left({-\frac{x^2}{2}}\right)
$$
and
$$
N(x)=\int_{-\infty}^{x}n(u) d u
$$

\noindent Following Ropper-Rutkowski (\cite{RR}), we can isolate the volatility $\sigma_N$ in the pricing formula.

\begin{defi}\label{defx}
Let us denote by ${\rm TV} (K, T )$ (or simply ${\rm TV}$) the time-value of a European call option struck at strike $K$ with maturity $T$: ${\rm TV} (T , K):=C(T,K)-(S-K)_+$.
\end{defi}

\begin{prop}\label{TVGamma}
In the Bachelier model,

\ba
\label{BSNG}
{\rm TV}(T,K) =
\left\lbrace
\begin{array}{l}
 \frac{{|S-K|}}{4 \sqrt{\pi}} \Gamma \left( -\frac{1}{2},\frac{(S-K)^2}{2 \sigma_N^2 T}\right)   \hspace{0.8cm} \textrm {if} \quad K\neq S \\
 \frac{\sigma_N \sqrt{T}}{\sqrt{2\pi}} \hspace{3.5cm} \textrm {otherwise} \\
\end{array}
\right.
\ea 
where $\Gamma \left(a,z\right)$ is the incomplete Gamma function:
$$    
\Gamma \left(a,z \right) = \displaystyle \int_z^{+\infty} u^{a-1}\exp(-u){\rm d} u
$$

\end{prop}

  The proof is given in the Appendix. It is clear from Proposition \ref{TVGamma} that the real-valued
  function $T\mapsto C (T,K)$ is non-decreasing, positive, $C(0,K)=(S-K)_+$ 
  and $\displaystyle\lim_{T\rightarrow +\infty} C(T,K)=+\infty$.  So, given the price of a European
  call option $C$, there is a unique real number $\sigma_N (T,K)$ such that
  $C(T,K)=C$ with a normal volatility  $\sigma_N=\sigma_N (T,K)$. We say that $\sigma_N (T,K)$ is the
  normal implied volatility. 

\begin{note}
We remark that $\displaystyle\frac{TV}{|K-S|}$ depends only of
$\displaystyle\frac{\sigma_N\sqrt{T}}{|S-K|}$ (only one variable).
\end{note}

  One of the interests of Proposition \ref{TVGamma}
  is that there are efficient algorithms to compute the
  inverse of the incomplete Gamma function. In particular, it is implemented
  in Matlab. Therefore, it is always easy to get the implied normal volatility
  from call prices (\cite{LiLee}) . Such a task is not always easy in the lognormal
  case (\cite{jaeckel}), especially when we are far from the money.

\begin{coro}\label{expanTV}
Let $p$ be an integer. Then,
$$
{\rm TV}(T,K) = \frac{(\sigma_N^2 T)^{\frac{3}{2}}}{\sqrt{2 \pi}(S-K)^2} \exp\left({-\frac{(S-K)^2}{2 \sigma_N^2 T}}\right) \left( \sum_{k=0}^{p-1}(-1)^{k}\frac{(2 k +1)!}{k!} \left( \frac{\sigma_N^2 T}{(S-K)^2}\right)^{k}+R_p\right)
$$

with

$$
|R_p|\leq \frac{(2 p + 1)!}{p!} \left( \frac{\sigma_N^2 T}{(S-K)^2}\right)^{p}
$$
\end{coro}

  This last equation comes naturally from a well known asymptotic expansion
  of $\Gamma(a,z)$ for large $z$ (see Formula 6.5.32 in (\cite{Abra})).

\begin{note}
    From either pricing formula (\ref{BSN}) or (\ref{BSNG}), we
    can notice that we can use the same demarch to price large strikes and short maturities
		European options (as expansions in both cases are similar)... This comes from the fact that:    
    
    $$
    C (\lambda^2 T,\lambda S + (1-\lambda) K) = \lambda C (T,K)
    $$
    for any non-negative real $\lambda$. This is particular to the Bachelier model.
\end{note}

	To compare with the lognormal case, it can be advantageous to introduce the following notations.
	
\begin{defi}
	For $K\not=S$, we set (the symbol N stands for ``Normal'')
	$\theta_N:=\displaystyle\frac{\sigma_N\,\sqrt{T}}{S},\,x_N:=\displaystyle\frac{K}{S}-1,\,
	\gamma_N:=\ln\left(\displaystyle\frac{4\,\sqrt{\pi}}{|x_N|}\right),\,u_N:=\displaystyle\frac{2\theta_N^2}{x_N^2},\,
	\gamma_N:=\ln\left(\displaystyle\frac{4\sqrt{\pi}}{|x_N|}\right),\, 
	\lambda:=-\displaystyle\frac{1}{\ln\left(\frac{{\rm 	TV}(T,\,K)}{S}\right)}$.
\end{defi}

	\noindent Then, by Corollary \ref{expanTV}, for $K\not= S$ and $p\in{\mathbb N}^*$,
	\begin{equation}\label{tvanalo}
	\displaystyle\frac{4\sqrt{\pi}}{|x_N|}
	\displaystyle\frac{{\rm TV}\,(T,K)}{S}=u_N^{\frac{3}{2}}\,{\rm e}^{-\frac{1}{u_{N}}}\,
	\left(\displaystyle\sum_{k=0}^{p-1}\,
	\displaystyle\frac{(-1)^k}{2^k}\,a_N^{(k)}\,u_N^k+R_N^{(p)}\right)
	\end{equation}
	with $R_N^{(p)}\in O\left(\theta_N^{2p}\right)$ and 
	$a_N^{(k)}=\displaystyle\frac{(2 k +1)!}{k!}$.
	
	\smallskip
	\noindent
	In comparison to the normal case, in the lognormal case, we have (\cite{grunspan}):
	
	\begin{equation}
	\label{tvln}
	\displaystyle\frac{4\sqrt{\pi}{\rm e}^{-\frac{x_{LN}}{2}}}{|x_{LN}|}
	\displaystyle\frac{{\rm TV}\,(T,K)}{S}=u_{LN}^{\frac{3}{2}}\,{\rm e}^{-\frac{1}{u_{LN}}}\,
	\left(\displaystyle\sum_{k=0}^{p-1}\,
	\displaystyle\frac{(-1)^k}{2^k}\,a_{LN}^{(k)}\,u_{LN}^k+R_{LN}^{(p)}\right)	
	\end{equation}
	with $u_{LN}:=\displaystyle\frac{2\theta_{LN}^2}{x_{LN}^2},
	\,\theta_{LN}:=\sigma_{LN}\sqrt{T},\,	x_{LN}:=\ln\left(\displaystyle\frac{K}{S}\right),\,
	R_{(LN)}^p\in O\left(\theta_{LN}^{2p}\right)$,
	$$
	a_{LN}^{(k)}:=(2k+1)!!\displaystyle\sum_{k=0}^{n}
	\displaystyle\frac{1}{j!\,(2j+1)!!}\,\left(\displaystyle\frac{x_{LN}^2}{8}\right)^{j}
	$$
	and $(2k+1)!!:=\displaystyle\prod_{j=0}^{k}(2j+1)$.	Here, $\sigma_{LN}$ denotes the lognormal implied volatility 
	and ``LN'' stands for ``Lognormal''.
	
\section{Asymptotics of the implied normal volatility}
  Let us assume that $K\neq S$. Using (\ref{tvanalo}), we get:
	\begin{equation}\label{davn}
	u_N^{\frac{3}{2}}{\rm e}^{-\frac{1}{u_N}}\left(\displaystyle\sum_{k=0}^{p-1}\,\alpha_k u_N^k+O\left(u_N^{p}\right)\right)
	={\rm e}^{\gamma_N}\,{\rm e}^{-\frac{1}{\lambda}}
	\end{equation}
	with $\alpha_k:=\displaystyle\frac{(-1)^k}{2^k}a_k$.	Therefore, by Lemma 1 of \cite{grunspan}, 
	we get the following 	proposition.
	
	\begin{prop}\label{laprop}
	Let us denote by ${\rm TV}$ the time-value of a European call option, $\sigma_{N}$ its implied normal volatility 
	and $T$ the maturity of the option. Set $\lambda:=-\displaystyle\frac{1}{\ln(\frac{{\rm TV}}{S})},
	\,\gamma_N:=\ln\left(\displaystyle\frac{4\sqrt{\pi}}{|x_N|}\right)$ and $x_N=\frac{K}{S}-1$. Let us assume that $K\neq S$.
	Then, in the case when $T\rightarrow 0$, we have the following expansion for the time-variance of the call option: 
	$\sigma_{N}^2 T=\displaystyle\frac{(S-K)^2}{2}\, u_N$ with
	
	\begin{equation}\label{kn}
	u_N = \lambda-\displaystyle\frac{3}{2}\lambda^2\ln \lambda
	+\gamma_N\lambda^2+\displaystyle\frac{9}{4}\lambda^3\ln^2(\lambda)+\left(\displaystyle\frac{9}{4}
	-3\gamma_N\right)\lambda^3\ln(\lambda)+\left(\gamma_N^2-\displaystyle\frac{3}{2}\gamma_N+
	\displaystyle\frac{3}{2}\right)\lambda^3
	+o\left(\lambda^3\right)
	\end{equation}
	\end{prop}
	
	\noindent In the lognormal case \cite{grunspan}, for short expiries, the asymptotic expansion of 
	$\sigma_{LN}^2 T$ is given by 
	$\sigma_{LN}^2 T=\displaystyle\frac{\ln^2(\frac{K}{S})}{2}\, u_{LN}$ with
	
	\begin{equation}\label{kln}
	u_{LN} = \lambda-\displaystyle\frac{3}{2}\lambda^2\ln \lambda
	+\gamma_{LN}\lambda^2+\displaystyle\frac{9}{4}\lambda^3\ln^2(\lambda)+\left(\displaystyle\frac{9}{4}
	-3\gamma_{LN}\right)\lambda^3\ln(\lambda)+\left(\gamma_{LN}^2-\displaystyle\frac{3}{2}\gamma_{LN}-\alpha'_1\right)\lambda^3
	+o\left(\lambda^3\right)
	\end{equation}
	
	\noindent where $\gamma_{LN}:=\ln\left(\displaystyle\frac{4\sqrt{\pi}{\rm e}^{-\frac{x_{LN}}{2}}}{|x_{LN}|}\right)$ 
	and $\alpha'_1:=-\displaystyle\frac{x_{LN}^2}{16}-\displaystyle\frac{3}{2}$.
	
	\medskip
	\noindent
	First, we note that $\lambda=u_{LN}+o(u_{LN})$. Then, comparing the two results (\ref{kn}) and (\ref{kln}) for
	$K\not= S$, we obtain:
	\ba
	\nonumber 
	u_{N}=u_{LN} + \left(\gamma_N-\gamma_{LN}\right)u_{LN}^2+O\left(u_{LN}^3\ln(u_{LN})\right)
	\ea
	So,
	\ba
	\sigma_N^2 =\left(\displaystyle\frac{S\,x_N}{x_{LN}}\right)^2\sigma_{LN}^2
	+2\left(\gamma_N-\gamma_{LN}\right)\displaystyle\frac{S^2\,x_N^2\,\sigma_{LN}^4}{x_{LN}^4}\, T
	+O\left(T^2\ln(T)\right)
	\ea
	Since $\displaystyle\frac{x_N}{x_{LN}}=\displaystyle\frac{S-K}{\ln S-\ln K}>0$, we deduce that
	\begin{eqnarray}
	\sigma_N&=&\displaystyle\frac{S\,x_N}{x_{LN}}\,\sigma_{LN}
	\left(1+2\left(\gamma_N-\gamma_{LN}\right)\displaystyle\frac{\sigma_{LN}^2}{x_{LN}^2}T\right)^{\frac{1}{2}}+O(T^2\ln T)\\
	&=&\displaystyle\frac{S\,x_N}{x_{LN}}\,\sigma_{LN}
	\left(1+\left(\gamma_N-\gamma_{LN}\right)\displaystyle\frac{\sigma_{LN}^2}{x_{LN}^2}T\right)+O(T^2\ln T)
	\end{eqnarray}
	Moreover, we have:
	\begin{equation}
	\nonumber
	\gamma_N-\gamma_{LN}=\displaystyle\frac{x_{LN}}{2}+\ln\left(\displaystyle\frac{x_{LN}}{x_{N}}\right)
	\end{equation}	 
	Hence, we get:
	\begin{equation}
	\nonumber
	\sigma_N=\displaystyle\frac{S-K}{\ln S-\ln K}\,\sigma_{LN}
	\left[1-\displaystyle\frac{\ln\left(\displaystyle\frac{1}{\sqrt{K S}}\displaystyle\frac{S-K}{\ln S-\ln K}\right)}
	{\left(\ln S-\ln K\right)^2}\,\sigma_{LN}^2T
	\right]+O(T^2\ln T)
	\end{equation}
	Note also that at the money, the situation is quite easy. On the one hand, we have (cf Proposition \ref{TVGamma})
	\begin{equation}
	\sigma_N=\sqrt{\displaystyle\frac{2\pi}{T}}\,C.	
	\end{equation}
	\nonumber
	On the other hand, we have (\cite{grunspan}, Proposition 2):
	\begin{equation}
	\nonumber
	C={\rm erf}\left(\displaystyle\frac{\sigma_{LN}\sqrt{T}}{2\sqrt{2}}\right)
	\end{equation}
	Therefore, we state the following result.

	\begin{coro}\label{maincoro}
	$\bullet$ If $K\not= S$, we have:
	\begin{equation}
	\label{fortankov}
	\sigma_N=\displaystyle\frac{S-K}{\ln S-\ln K}\,\sigma_{LN}
	\left[1-\displaystyle\frac{\ln\left(\displaystyle\frac{1}{\sqrt{K S}}\,\displaystyle\frac{S-K}{\ln S-\ln K}\right)}
	{\left(\ln S-\ln K\right)^2}\,\sigma_{LN}^2T
	\right]+O(T^2\ln T)
	\end{equation}
	In particular, $\sigma_N\sim\displaystyle\frac{S-K}{\ln S-\ln K}\,\sigma_{LN}$ when $T\rightarrow 0$.
	
	\smallskip
	\noindent
	$\bullet$
	At the money, we have $\sigma_N=\sqrt{\displaystyle\frac{2\pi}{T}}\,S\,
	{\rm erf}\left(\displaystyle\frac{\sigma_{LN}\sqrt{T}}{2\sqrt{2}}\right)$
	and $\sigma_N=S\,\sigma_{LN}\left(1-\displaystyle\frac{\sigma_{LN}^2 T}{24}\right)+o(T)$.
	In particular, $\sigma_N\sim S\sigma_{LN}$ when $T\rightarrow 0$.
	\end{coro}
	
	\begin{note}
	When $K\rightarrow S$, we can check that 
	$\displaystyle\frac{\ln\left(\displaystyle\frac{1}{\sqrt{K S}}\,\displaystyle\frac{S-K}{\ln S-\ln K}\right)}
	{\left(\ln S-\ln K\right)^2}\longrightarrow \displaystyle\frac{1}{24}$.
	\end{note}
	
	\noindent
	In other terms, for $K\not= S$,
	\begin{equation}
	\label{main}
	\sigma_{LN} = \frac{1}{S} \frac{\ln m}{m-1} \sigma_N
	\end{equation}
	where $m=\displaystyle\frac{K}{S}=x_N+1$ is the moneyness.
	
	This formula was known (even if it was not stated explicitly) in the SABR model 
	(see the Hagan et al formula \cite{hagan}). By differentiating Formula (\ref{main}) 
	with respect to $m$, it turns out that the Black-Scholes skew $\frac{\partial \sigma_{LN}}{\partial m}$ 
	at the money ($m=1$) generated by the Bachelier model is $\frac{\partial\sigma_{LN}}{\partial m} 
	= -\frac{1}{2} \frac{\sigma_N}{S}$ ($\sigma_{LN}$ is by definition the implied lognormal volatility).
	Therefore, the Bachelier model is highly skewed ATM (a slope of $-50\%\times \frac{\sigma_N}{S}$). Another way to 
	explain this feature is that given call prices, when we use the BS model, the function $\sigma_{LN}$ is a decreasing 
	and convex function of $m$, i.e., it generates a skew,  while  the function $\sigma_{N}$ is a rather flat 
	function of $m$. Thus, normal volatility is most suited for products such as swaptions for instance.

	\section{Comparing greeks and delta-hedged portfolios}
	Let us denote by $\Delta_N, \Gamma_N, \nu_N, \Theta_N$ (resp. $\Delta_{LN}, \Gamma_{LN}, \nu_{LN}, \Theta_{LN}$), the delta, gamma, vega and theta in the Bachelier (resp. Black-Scholes) model. For instance, $\nu_N=\frac{\partial C}{\partial \sigma_N}.$ By differentiating (\ref{BSN}), we get:
	\ba\label{DeltaN}
	\Delta_N = N\left(\frac{S-K}{\sigma_{N}\sqrt{T}}\right)
	\ea
	On the other hand, it is known that:
	\ba
	\Delta_{LN} = N\left(\frac{\ln S - \ln K}{\sigma_{LN}\sqrt{T}}+\frac{1}{2}\sigma_{LN}\sqrt{T} \right)
	\ea
	So, by Corollary \ref{maincoro}, we get: $\Delta_{LN}\sim\Delta_{N}$ for a maturity $T\ll 1$. 
	By differentiating (\ref{DeltaN}), we obtain:
	\ba\label{GammaN}
	\Gamma_N = \frac{1}{\sigma_{N}\sqrt{T}} \hspace{0.1cm} n\left(\frac{S-K}{\sigma_{N}\sqrt{T}}\right)
	\ea
	In the Black-Scholes model, we have:rrrrrr
	\ba
	\Gamma_{LN} = \frac{1}{S\sigma_{LN}\sqrt{T}} \hspace{0.1cm} n
	\left(\frac{\ln S - \ln K}{\sigma_{LN}\sqrt{T}}+\frac{1}{2}\sigma_{LN}\sqrt{T}\right)
	\ea
	Hence, with the help of Corollary \ref{maincoro},
	\ba
	\Gamma_{N} \sim S \frac{\ln S - \ln K}{S-K}\Gamma_{LN}
	\ea
	Now, we consider the Vega. It is shown in the Appendix (Formula (\ref{fortheta})) that:
	\ba
	C (T,K) = (S-K)_+ + S \displaystyle\int_{0}^{\frac{\sigma_N\sqrt{T}}{S}} n\left(\frac{1-\frac{K}{S}}{u}\right) {\rm d} u
	\ea
	So, differentiating by $\sigma_N$, we get:
	\ba
	\nu_N = S \sqrt{\frac{T}{2 \pi}}\exp\left(\frac{(S-K)^2}{2 \sigma_N^2 T}\right)
	\ea
	In contrast, the vega in the Black-Scholes model is 
	\ba
	\nu_{LN} = S \sqrt{\frac{T}{2 \pi}}\exp\left(\frac{(\ln S - \ln K + \frac{1}{2} \sigma_{LN}^2 T)^2}{2 \sigma_{LN}^2 T}\right)
	\ea
	
	In the same way, we can compare the two thetas. So, we get the following proposition.
	
	\begin{prop}\label{GreeksNLN}
	When $T\rightarrow 0$, and under the hypothesis of bounded volatilities, we have:
	
	\begin{eqnarray}
	\Delta_N&\sim&\Delta_{LN}\\
	\nu_N&\sim&\nu_{LN}\\
	\Gamma_N&\sim&S \frac{\ln S - \ln K}{S-K}\Gamma_{LN}\\
	\Theta_N&\sim&\Theta_{LN}\frac{S-K}{S\left(\ln S - \ln K\right)} \Theta_{LN}
	\end{eqnarray}
	\end{prop}
	
	The first equivalence $\Delta_N\sim\Delta_{LN}$ shows that hedging in the Bachelier framework is more or less like hedging in a Black-Scholes framework. However, the ``breakeven move'' of a delta-hedged portfolio is not the same. By definition, the ``breakeven move'' of a delta-hedged portfolio is the number $\mu$ such that over a short horizon $\delta t$, ${\rm P\& L} > 0$ if the change in $S$ is $>\mu$. In general, we have:
	
	\begin{eqnarray}
	{\rm P\& L}&=&-\Theta \delta t + \frac{1}{2}\Gamma(\Delta S)^2\\
	&=&\frac{1}{2}\Gamma\left[(\Delta S)^2-\mu^2\right]
	\end{eqnarray}
	
	So, with $\delta t = 1$, 
	
	\ba
	\mu = \sqrt{\frac{2 \Theta}{\Gamma}}.
	\ea
	
	Using Proposition \ref{GreeksNLN}, we get that the ``breakeven move'' $\mu_{LN}$ in the Black-Scholes model is related with the ``breakeven move'' $\mu_N$ in the Bachelier model by:
	\ba
	\mu_{LN} = \frac{\ln m}{m-1} \mu_N
	\ea
	with $m=\frac{K}{S}$. So, at the money, $\mu_{LN}\sim\mu_N$. However, if $K<S$ (resp. $K>S$) then $\mu_{LN}>\mu_N$ (resp.
	$\mu_{N}>\mu_{LN}$). We have represented in Fig 1, the function $m\mapsto\displaystyle\frac{\ln(m)}{m-1}$ which gives the smile of the Bachelier model (cf Corollary \ref{maincoro}) as well as the ratio of ``breakeven moves'' $\displaystyle\frac{\mu_{LN}}{\mu_{N}}$ .

	So, depending on the view of the trader on the short term dynamic of the underlying (normal or lognormal diffusion), he will adjust or not the ``breakeven move'' of its delta-hedged portfolio by the factor $\frac{\ln m}{m-1}$.
	
	\begin{figure}
	\centering \includegraphics[width=10cm,height=8cm]{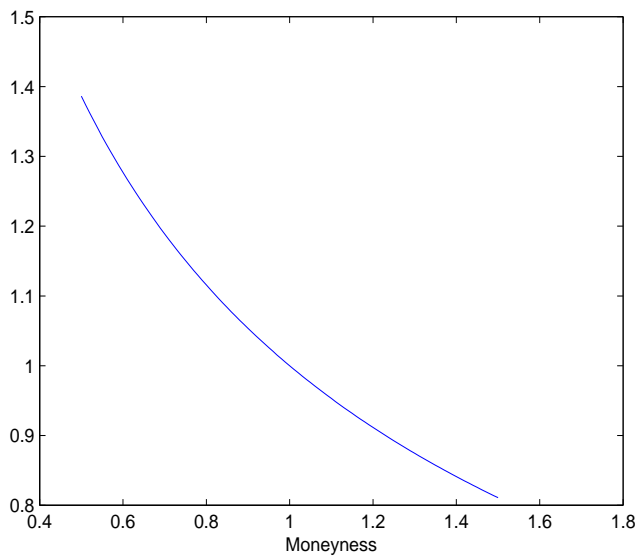}
	\caption{The smile of volatility of the Bachelier model (normalized by the factor $\displaystyle\frac{\sigma_N}{S}$ where $\sigma_N$ denotes the normal volatility) and the ratio of ``breakeven moves'' $\mu_{LN}/\mu_N$.}\label{fig:F1}
	\end{figure}

	\section*{Appendix}
	{\bf Proof of Proposition \ref{TVGamma}.}
	We have:
	
	\ba
	\frac{C (T,K)}{S} = f (\xi,\theta) 
	\ea
	
	with $\xi:=\frac{K}{S}$ and $\theta:=\frac{\sigma_N \sqrt{T}}{S}$ and
	
	\ba
	f (\xi,u) := (1-\xi) N \left(\frac{1-\xi}{u}\right)+ u n \left(\frac{1-\xi}{u}\right)
	\ea
	
	We have:
	
	\ba
	\begin{array}{rcl}
	\displaystyle{\frac{\partial{f}}{\partial u} (\xi,u)} & := & \displaystyle{-\frac{(1-\xi)^2}{u^2}\, 
	n \left(\frac{1-\xi}{u}\right)+  n \left(\frac{1-\xi}{u}\right) + u\left(-\frac{1-\xi}{u^2}\right).
	\left(-\frac{1-\xi}{u}\right) n \left(\frac{1-\xi}{u}\right)}\\
	& = & \displaystyle n \left(\frac{1-\xi}{u}\right)
	\end{array}
	\ea
	
	We have used: $n' (\xi) = -\xi\, n(\xi)$. Since $f (\xi,0) = \frac{C (0,K)}{S} = (1-\xi)_+$, we deduce that:
	
	\ba
	f (\xi,\theta) = (1-\xi)_+ + \displaystyle\int_{0}^{\theta} n\left(\frac{1-\xi}{u}\right) du
	\ea
	
	Set
	
	\ba
	F (\xi,\theta) = \displaystyle\int_{0}^{\theta} n\left(\frac{1-\xi}{u}\right) du
	\ea
	
	Then,
	
	\ba\label{fortheta}
	\frac{C (T,K)}{S} = (1-\xi)_+ + F \left(\frac{K}{S},\frac{\sigma \sqrt{T}}{S}\right)
	\ea
	 Let us assume that both $\theta\neq 0$ and $\xi\neq 1$. With the change of variable $v=\frac{1-\xi}{u}$, we get:
	 
	\ba
	F (\xi,\theta) = |1-\xi| \displaystyle\int_{\frac{|1-\xi|}{\theta}}^{+\infty} \frac{n (v)}{v^2} dv
	\ea
	
	So, with the new change of variable $u=\frac{1}{2}v^2$, we have:
	\ba
	\begin{array}{rcl}
	F (\xi,\theta) & = & \displaystyle\frac{1}{4 \sqrt{\pi}}\, |1-\xi|\,
	\displaystyle\int_{\frac{|1-\xi|^2}{2\theta^2}}^{+\infty} u^{-\frac{3}{2}} \exp(-u) du\\
	&=&\displaystyle\frac{1}{4 \sqrt{\pi}}\, |1-\xi|\, \Gamma\left(-\frac{1}{2},\frac{|1-\xi|^2}{2\theta^2}\right)
	\end{array}
	\ea
	
	where $\Gamma (a,z)$ is the incomplete Gamma function. At the money, we have simply:
	
	\ba
	C (T,K) = \frac{\sigma_N \sqrt{T}}{\sqrt{2 \pi}}.
	\ea
	
	\end{document}